\documentclass[useAMS,usenatbib,twocolumn]{mnras}
\usepackage{graphicx}
\usepackage{times}
\usepackage{amssymb}
\usepackage{amsmath}
\usepackage{multirow, makecell}
\usepackage{subfig}
\usepackage{hyphenat}
\usepackage{hyperref}
\usepackage{sansmath}
\usepackage{tabularx}
\begin{document}

\title[Standard Siren Speeds] 
{Standard Siren Speeds: Improving velocities in gravitational-wave measurements of $H_{0}$.}

\author[C. Howlett et. al.]{\parbox{\textwidth}{
Cullan Howlett\thanks{Email: c.howlett@uq.edu.au}$^{1}$,
Tamara M. Davis$^{1}$
}
  \vspace*{4pt} \\ 
$^{1}$ School of Mathematics and Physics, The University of Queensland, Brisbane, QLD 4072, Australia. \\
}

\pagerange{\pageref{firstpage}--\pageref{lastpage}} \pubyear{2019}
\maketitle
\label{firstpage}

\begin{abstract}
We re-analyze data from the gravitational-wave event GW170817 and its host galaxy NGC4993 to demonstrate the importance of accurate total and peculiar velocities when measuring the Hubble constant using this nearby Standard Siren. We show that a number of reasonable choices can be made to estimate the velocities for this event, but that systematic differences remain between these measurements depending on the data used. This leads to significant changes in the Hubble constant inferred from GW170817. We present Bayesian Model Averaging as one way to account for these differences, and obtain $H_{0}=66.8^{+13.4}_{-9.2}\,\mathrm{km\,s^{-1}\,Mpc^{-1}}$. Adding additional information on the viewing angle from high resolution imaging of the radio counterpart refines this to $H_{0}=64.8^{+7.3}_{-7.2}\,\mathrm{km\,s^{-1}\,Mpc^{-1}}$. During this analysis we also present an alternative Bayesian model for the posterior on $H_{0}$ from Standard Sirens that works more closely with observed quantities from redshift and peculiar velocity surveys. Our results more accurately capture the true uncertainty on the total and peculiar velocities of NGC4993 and show that exploring how well different datasets characterize galaxy groups and the velocity field in the local Universe could improve this measurement further.  These considerations impact any low-redshift distance measurement, and the improvements we suggest here can also be applied to standard candles like type Ia supernovae.  GW170817 is particularly sensitive to peculiar velocity uncertainties because it is so close.  For future standard siren measurements the importance of this error will decrease as (a) we will measure more distant standard sirens, and (b) the random direction of peculiar velocities will average out with more detections.
\end{abstract}


\section{Introduction}
On the $17^{\mathrm{th}}$ August 2017 the Advanced LIGO \citep{LIGO2015} and Advanced Virgo \citep{Acernese2015} detectors observed gravitational waves originating from event GW170817. Modelling of this signal later identified this event as a result of the merger of two compact neutron stars \citep{Abbott2017a, Abbott2019}. In the short time after, electromagnetic counterparts were detected across a number of wavelengths \citep{Abbott2017b} resulting in the first multi-messenger detection of a gravitational wave.

\begin{table*}
\setlength{\extrarowheight}{3pt}
\caption{Definitions of redshifts and their corresponding velocities used in this work, the relationship between them, and the physical description of their origin.}
\centering
\resizebox{\textwidth}{!}{
\begin{tabular}{lclcp{9.2cm}} \hline
\multicolumn{2}{c}{Redshifts} & \multicolumn{2}{c}{Velocities} & \multirow{2}{*}{Description}  \\ 
Name & Symbol & Name & Symbol &  \\ \hline
Observed redshift & $z_{\mathrm{obs}}$ & Total velocity & $v_{t}$ & The redshift/velocity of an object as measured by us in the heliocentric (Sun-centred) frame, i.e., after correcting for the rotation of the Earth, and its motion around the Sun, \textit{but without} any corrections for the Sun's motion relative to the CMB dipole, or correction for the observed object's motion. Equal to the combined contributions from the Sun's peculiar, object's peculiar, and recession velocities, i.e., the product of the relevant $(1+z)$ factors, Eq~\ref{eq:redshiftproper}. \\
Cosmological redshift & $\bar{z}$ & Recession velocity & $v_{r}$ & The redshift/velocity of the observed object due to the expansion of the Universe.  \\
Object's peculiar redshift & $z^{\mathrm{obj}}_{\mathrm{p}}$ & Object's peculiar velocity & $v^{\mathrm{obj}}_{\mathrm{p}}$ & The component of the redshift/velocity of the observed object due to its motion towards/away from us, in departure from the expansion of the universe. The peculiar redshift and peculiar velocity are related by Eq.~\ref{eq:redvel}. \\
Sun's peculiar redshift & $z^{\mathrm{Sun}}_{\mathrm{p}}$ & Sun's peculiar velocity & $v^{\mathrm{Sun}}_{\mathrm{p}}$ & The component of the redshift/velocity of an object due to the motion of \textit{our} Solar system in the direction of the observed object, relative to the CMB rest frame. This is typically inferred from measurements of the amplitude and direction of the CMB dipole. The Sun's peculiar redshift and peculiar velocity are related by Eq.~\ref{eq:redvel}. \\
CMB-frame redshift & $z_{\mathrm{cmb}}$ & CMB-frame velocity & $v_{\mathrm{cmb}}$ & The redshift/velocity of the object after correcting for our Solar system's peculiar motion with respect to the CMB dipole. This should represent the redshift with us in the comoving frame, but may still have contributions from the observed galaxy's/host's peculiar velocity. This should be computed using $(1 + z_{\mathrm{cmb}}) = \frac{(1+z_{\mathrm{obs}})}{(1+z^{\mathrm{Sun}}_{\mathrm{p}})}$ but is often approximated. \vspace{3pt} \\ \hline
\end{tabular}}
\label{tab:nomenclature}
\end{table*}

The presence of the electromagnetic counterparts gave a precise determination (with probability of chance association $P<0.004\%$; \citealt{Abbott2017c}) that the host of the gravitational wave was NGC4993, a low redshift lenticular galaxy. The combination of gravitational wave and host identification then allowed for the first ever ``Standard Siren" measurement of the Hubble constant, $H_{0}$, analogous to methods using ``Standard Candles" (type Ia supernovae - SN Ia; Cepheid variable stars, etc.,) and ``Standard Rulers" (e.g., Baryon Acoustic Oscillations - BAO; the Cosmic Microwave Background - CMB). This determination of the Hubble constant was made possible through the combination of cosmological luminosity distance inferred from the gravitational-wave signal, combined with the observed total and peculiar velocities of NGC4993. The luminosity distance inferred from the gravitational wave is consistent with independent measurements of the distance to NGC4993 \citep[e.g.][]{Hjorth2017,Im2017,Cantiello2018}. It is worth noting that in cosmological terms, Standard Sirens and Candles are very similar; both measure the luminosity distance through calibration of an astrophysical signal, which in the case of Standard Sirens is the frequency and rate of change of frequency of the gravitational wave. This can be compared to Standard Rulers, which measure a different cosmological quantity, the angular diameter distance. 

The Hubble constant is one of the fundamental constants describing our cosmological model. It describes how fast the Universe is expanding, and how fast objects are receding from each other. Precise determination of this constant has been one of the foremost goals of cosmology since its discovery, with the majority of measurements using either Standard Candles or Standard Rulers. In recent years, tensions have arisen between measurements from these two methods. Results from a combination of Planck CMB and various BAO measurements prefer $H_{0}=67.66 \pm 0.42\,\mathrm{km\,s^{-1}\,Mpc^{-1}}$ \citep{Planck2018}, however this requires assuming a $\Lambda$CDM cosmological model to extrapolate the constraints from high redshift. Results using the local distance ladder (SNe anchored using Cepheids and local geometric distances) prefer $H_{0}=74.03 \pm 1.42\,\mathrm{km\,s^{-1}\,Mpc^{-1}}$ \citep{Riess2019}.\footnote{However, very recent results suggest that changing the Cepheid anchor to Tip of the Red Giant Branch (TRGB) stars can cause significant shifts in the preferred Hubble constant \citep{Freedman2019}.} The tension between these two is currently at the level of $\sim4.5\sigma$ and hints at the presence of unknown systematics or new fundamental physics. It seems unlikely that this will be resolved over the coming years without additional, independent measurements. Standard Sirens using gravitational waves present the exciting prospect of such a measurement, and may identify a way to resolve the current tension.

Standard Sirens are one of the cleanest distance measurements available, but they are not without their own sources of uncertainty, both statistical and systematic. Especially in the infancy of this technique, with few measurements available, these could cause biases in the recovered constraints on the Hubble constant.  The potential source of systematic uncertainty we concentrate on in this paper is the influence of peculiar velocities.  Peculiar velocities have long been know to potentially bias measurements of $H_0$, particularly if the sources are nearby \citep[e.g.][]{Dressler1987,Sandage1990,Tonry2000,Tully2008}. We begin by summarising how peculiar velocities impact $H_0$ estimates in general, and then consider the specifics of the peculiar velocity estimates of GW170817's host, NGC4993.

With determination of the host galaxy and a measurement of the luminosity distance $d_{L}$, we can infer the Hubble constant via
\begin{equation}
(1 + z_{\mathrm{obs}}) = (1+\bar{z}(d_{L}, H_{0}, z_{\mathrm{obs}}))(1 + z^{\mathrm{obj}}_{\mathrm{p}})(1+z^{\mathrm{Sun}}_{\mathrm{p}}),
\label{eq:redshiftproper}
\end{equation}
where $z_{\mathrm{obs}}$ is the observed redshift and $z^{\mathrm{obj}}_{\mathrm{p}}$ is the peculiar redshift that arises from the peculiar velocity of the observed object.  $\bar{z}(d_{L}, H_{0}, z_{\mathrm{obs}})$ is the cosmological redshift of the object, which can be computed given a choice of the Hubble constant and an inference of the luminosity distance.  $z^{\mathrm{Sun}}_{\mathrm{p}}$ is an additional redshift arising from our Sun's motion with respect to the comoving frame, typically calculated using the CMB dipole. For reference, the relationship between peculiar redshift (of either an observed object, or our Solar system) and velocity is given by
\begin{equation}
1+z =\sqrt{\frac{1+v/c}{1-v/c}} \approx 1+v/c
\label{eq:redvel}
\end{equation}
where $c$ is the speed of light and the approximation holds as long as $v_{p} \ll c$ \citep{Davis2019}. Note, this relationship is only appropriate for objects moving within their local inertial frame and so is not appropriate for converting the cosmological redshift $\bar{z}$ to a velocity \citep{Davis2004}.

It is common to see Eq.~\ref{eq:redshiftproper} computed using a number of approximations, where first the CMB dipole is used to convert the observed redshift to the redshift we would have observed if we had no peculiar velocity and were comoving observers (in the CMB frame) using the approximation $z_{\mathrm{cmb}} = z_{\mathrm{obs}} - z^{\mathrm{Sun}}_{\mathrm{p}}$ and then the Hubble constant is inferred using 
\begin{equation}
cz_{\mathrm{cmb}} = v^{\mathrm{obj}}_{\mathrm{p}} + H_{0}^{\mathrm{approx}}d_{L}/(1+z_{\mathrm{cmb}}).
\label{eq:redshiftapprox}
\end{equation}
However, we will tacitly avoid such approximations in this work and demonstrate that they are unnecessary (and inadequate) for inferring the Hubble constant from Standard Sirens. It is also common to use redshifts and velocities interchangeably, which leads to potential confusion/mistakes. For clarity, we provide in Table~\ref{tab:nomenclature} a list of definitions for the various terms used in this work.

Regardless of the approximations used, constraining the Hubble constant using gravitational waves (or indeed any local distance measurement) requires knowledge of the observed object's total and peculiar velocities. There are a number of methods to estimate these in combination. Firstly, the total velocity of the host can be measured spectroscopically and combined with a measurement of the peculiar velocity of the host. However, in this case, the total and peculiar velocities of the host galaxy are influenced by its motion within its local group or cluster, which includes non-linear effects from growth of structure and/or virialised motions that may be hard to account for in the peculiar velocity estimate. The peculiar velocity for a single object is also subject to considerable statistical error. It is often preferable, as was done in \cite{Abbott2017c}, to use instead the total and peculiar velocity of the group to which the object belongs such that non-linear motions are `smoothed-out' and the uncertainty in the peculiar velocity reduced. In this case, there still remains a choice of how to compute which galaxies belong to the same group as the host, and how to obtain the peculiar velocity of the group.

In terms of estimating the group peculiar velocity, there are two approaches commonly adopted. The first is to use a `peculiar velocity survey'; a catalogue of measured peculiar velocities estimated for objects with secondary distance measurements (more detail on these is given in Section~\ref{sec:pecvel}). Different interpolation methods can then be applied to this catalogue to estimate the peculiar velocity at the location of the group. In this case however, it is worth noting that peculiar velocity is not actually the directly observed quantity from peculiar velocity surveys (it is the change in magnitude or size caused by the underlying peculiar velocity) and hence the uncertainties in the measured peculiar velocities are not typically Gaussian distributed (unless an approximate estimator is used; \citealt{Watkins2015}).  The second method is to use measurements of the density field from galaxy redshift surveys combined with theoretical (linear or quasi-linear) predictions for the relationship between the density and velocity fields to reconstruct the gravitational infall onto large scale structures. Overall, this is to say that there are a number of choices one can make for how to estimate the joint total and peculiar velocities of the observed object. Should we use the properties of the host \textit{or} its group? If the latter, which group catalogue should we believe? Should the peculiar velocity be estimated using peculiar velocity surveys or reconstructions? 

\cite{Abbott2017c} present one such combination of methods to estimate both the CMB-frame redshift (which they treat as the total velocity) and peculiar velocity of the group containing NGC4993. These are given values of  $cz_{\mathrm{cmb}} = 3327\pm72\,\mathrm{km\,s^{-1}}$ and $v^{\mathrm{host}}_{\mathrm{p}}=310 \pm 150\,\mathrm{km\,s^{-1}}$ respectively in their canonical analysis. Using these values and Eq.~\ref{eq:redshiftapprox} they obtain the first ever measurement of $H_{0}$ using Standard Sirens, $H_{0}=70.0^{+12.0}_{-8.0}\,\mathrm{km\,s^{-1}\,Mpc^{-1}}$. This was improved upon by \cite{Hotokezaka2019} to $H_{0}=68.9^{+4.7}_{-4.6}\,\mathrm{km\,s^{-1}\,Mpc^{-1}}$ using Very Long Baseline Interferometer (VLBI) measurements of the centroid motion and afterglow lightcurve of the jet associated with GW170817, which substantially improves the degeneracy between the luminosity distance and observing angle.

In this work we will demonstrate that for the sole case of GW170817 and its host NGC4993 there are a number of alternative methods that could be used to obtain the total and peculiar velocities required for the Standard Siren determination of $H_{0}$. Although this is not an exhaustive list (for instance, we focus on peculiar velocity surveys, and don't consider the range of reconstructions one could also use), among these cases we identify systematic differences that are larger than the uncertainties commonly used for these measurements suggest. These differences translate into a range of Hubble constants; again wider than the uncertainty on $H_{0}$ from \cite{Abbott2017c} or \cite{Hotokezaka2019}. Although the error is currently large, studies (e.g., \citealt{Chen2018,Mortlock2018,Shafieloo2018,Zhang2019}) show that with only a handful of similar cases, we could obtain constraints on $H_{0}$ comparable to those from Standard Candles or Rulers. Whilst the impact of peculiar velocity uncertainties may reduce as more objects are detected at larger distances, it will be some time before peculiar velocities become a sub-dominant source of uncertainty. Hence, understanding and accounting for systematic differences in our estimates of the host's velocity is important if Standard Sirens are to be considered a reliable way to measure $H_{0}$ in the near-future. 

\begin{table*}
\setlength{\extrarowheight}{3pt}
\caption{Properties of groups containing NGC4993 from various group catalogues. The first row is the properties of NGC4993 itself. The second row contains the properties used in the original \protect\cite{Abbott2017c} analysis.  For each group we list the group identifier, the number of members $N$, the mean observed redshift and CMB-frame redshift (both multiplied by the speed-of-light to give units of $\mathrm{km\,s^{-1}}$) with error computed from the velocity dispersion divided by the square root of the number of members. We then compute the 3-D (not projected) distance from NGC4993 to the centre of the group assuming our fiducial cosmology to convert to Cartesian coordinates. We additionally identify how many members in the group have distance measurements in the Cosmic-Flows III compilation $N(v_{p})$, and compute the mean log-distance ratio and peculiar velocity. See Section~\ref{sec:pecvel} for a description of how these are computed. Each of the group catalogues can be obtained from its respective reference, except the catalogue based on the 6-degree Field Galaxy Redshift Survey (6dFGRS), which was obtained from A. Merson, D. Heath Jones and M. Colless (private communication). \protect\cite{Kourkchi2017} (Trimmed) indicates we have removed the galaxies from this group that were identified by \protect\cite{Hjorth2017} as only being loosely associated with the group.}
\centering
\resizebox{\textwidth}{!}{
\begin{tabular}{lccccccccc} \hline
\multirow{2}{*}{Reference} & \multirow{2}{*}{Group Name} & \multirow{2}{*}{$N$} &  Observed Redshift & CMB-frame Redshift & Velocity Dispersion & Distance from centre & \multirow{2}{*}{$N(v_{p})$} & Log-distance Ratio & Peculiar Velocity \\ 
& & & $cz_{\mathrm{obs}}$ ($\mathrm{km\,s^{-1}}$) & $cz_{\mathrm{cmb}}$ ($\mathrm{km\,s^{-1}}$) & $\sigma_{v}$ ($\mathrm{km\,s^{-1}}$) & ($\mathrm{h^{-1}\,Mpc}$) & & $\eta$ & $\langle v^{\mathrm{obj}}_{\mathrm{p}} \rangle$ ($\mathrm{km\,s^{-1}}$) \\ \hline
NGC4993 & - & 1 & $2916 \pm 15$ & $3228 \pm 15$ & - & - & - & - & -  \\
\cite{Abbott2017c} & - & 5 & - & $3327 \pm 72$ & - & - & - & - & $310 \pm 150$ \\ \hline 
\cite{Crook2008} (LDC) & 955 & 46 & $2558 \pm 72$ & $2878 \pm 72$ & 487 & $4.93 \pm 0.53$ & 18 & $0.0665 \pm 0.0150$ & $440 \pm 99$ \\
\cite{Crook2008} (HDC) & 763 & 5 & $3016 \pm 72$ & $3327 \pm 72$ & 160 & $1.13\pm 0.63$ & 1 & $0.0762 \pm 0.1000$ & $582 \pm 764$ \\
\cite{Lavaux2011} & 1338 & 10 & $3026 \pm 53$ & $3338 \pm 53$ & 169 & $1.19 \pm 0.52$ & 2 & $0.0676 \pm 0.0514$ & $518 \pm 394$ \\
\cite{Makarov2011} & NGC4993 & 15 & $2935 \pm 19$ & $3248 \pm 19$ & 72 & $0.41 \pm 0.14$ & 1 & $0.0645 \pm 0.0600$ & $481 \pm 448$ \\
\cite{Tully2015} & 100214 & 8 & $2997 \pm 51$ &  $3310 \pm 51$ & 143 & $1.19 \pm 0.51$ & 3 & $0.1136 \pm 0.0391$ & $864 \pm 297$ \\
\cite{Kourkchi2017} & 45466 & 22 & $2995 \pm 32$ & $3308 \pm 32$ & 151 & $0.87 \pm 0.35$ & 2 & $0.0676 \pm 0.0514$ & $514 \pm 391$ \\
\cite{Kourkchi2017} (Trimmed) & 45466 & 17 & $2919 \pm 13$ & $3231 \pm 13$ & 52 & $0.43 \pm 0.12$ & 1 & $0.0645 \pm 0.0600$ & $479 \pm 445$ \\
6dFGRS Groups & GRP0056 & 11 & $3028 \pm 51$ & $3341 \pm 51$ & 166 & $1.10 \pm 0.48$ & 1 & $0.0762 \pm 0.1000$ & $585 \pm 767$ \\ \hline
\end{tabular}}
\label{tab:recvel}
\end{table*}

It is worth noting that this problem is most prevalent for current Standard Siren measurements with identifiable hosts. In the case where the true host is unknown and marginalised over (e.g., \citealt{Fishbach2019, Soares-Santos2019}), the impact of peculiar velocity errors will be smaller, both because the uncertainty on  $H_{0}$ will be larger and because the peculiar velocities should be randomly oriented and will partially cancel out (although one would ideally hope for a small localisation area, which would mean the velocities will be correlated and cancel out less effectively).  Even at nearby distances with electromagnetic counterparts, the error in peculiar velocities should partially cancel out as more Standard Siren are detected.  However, we will need hundreds of Standard Sirens before this is effective and there could still be a residual bias even after many hundreds have been averaged if the peculiar velocity errors are asymmetric or systematic. These errors would also impact supernova measurements but as SNe are typically at higher observed redshift the impact is smaller. Cosmological studies usually reject any SNe closer than $z_{\mathrm{obs}} < 0.02$ which is where the typical peculiar velocity should contribute less than $5\%$ uncertainty to the total velocity.

We order this work as follows; In Section~\ref{sec:recvel} we present a number of determinations for the group total velocity (or rather the group observed redshift) of NGC4993. In Section~\ref{sec:pecvel} we do the same for the peculiar velocity/redshift. In both cases we identify a number of reasonable choices that give significantly different results. In Section~\ref{sec:hubble} we show how these choices change the inferred Hubble constant, but also offer a way in which such systematic differences could be folded into the marginalised constraints. We conclude in Section~\ref{sec:conclusion}. Where necessary, or unless otherwise stated, we assume a flat $\Lambda$CDM cosmology with $\Omega_{m}=0.3$. Given the low redshift of the sample, we do not expect changing this to affect our results (in particular any distance calculations), and any distances used herein that may affect the extraction or interpretation of the Hubble constant are given independent of its value (i.e., in units of $h^{-1}\,\mathrm{Mpc}$).

\section{Estimates of the Group observed redshift} \label{sec:recvel}
The galaxy NGC4993 has been identified as belonging to a larger group of galaxies in a number of different studies. The total and peculiar velocities of the group can be used in place of the individual galaxy in order to reduce the effects of non-linearities and internal peculiar motions and to improve the precision with which the peculiar velocity is known. The argument is as follows: the size of the cluster is small enough that the distance to the cluster is approximately the distance to the galaxy that hosted the gravitational wave.  For an object such as NGC4993 (with CMB-frame redshift $cz_{\mathrm{cmb}} = 3228 \pm 15\,\mathrm{km\,s^{-1}}$) the peculiar motion of the galaxy within its group or cluster could easily be as much as $20\%$ of the CMB-frame velocity. Hence, the internal peculiar motion is expected to give a larger error than the error due to using the centre of the cluster as the distance (see Table~\ref{tab:recvel} for estimates of such a distance for NGC4993). Similarly, using the redshift of the cluster is closer to the cosmological redshift than a galaxy within the cluster will be. 

A second benefit of using the velocities for the group is that the peculiar velocity of the group can be measured with higher precision than for an individual object. For peculiar velocities estimated using a peculiar velocity catalogue, the measurements for individual group members can be averaged over. For estimates obtained from reconstructions, the density field can be smoothed on scales comparable to the group size, such that linear theory can be more reliably used to estimate the velocities without need for additional uncertainties to account for non-linearities.

As a caveat, one downside of using the group velocities is the potential to mistake the host as a member of a group when it is not. Ensuring the group catalogues used to identify the host's group are robust is critical in obtaining reliable $H_{0}$ estimates. Comparing or marginalising over different group catalogues reduces the potential for these ``catastrophic errors'' somewhat, but may not mitigate against it entirely because many group catalogues use overlapping data or similar algorithms.

\cite{Abbott2017c} used the High Density Catalogue (HDC) of \cite{Crook2007,Crook2008} (which contains 5 members including NGC4993) to estimate a CMB-frame redshift of the group, after correcting for our own motion with respect to the CMB, of $cz_{\mathrm{cmb}} = 3327\pm72\,\mathrm{km\,s^{-1}}$. The corresponding observed redshift, without correction for our motion, taken directly from the group catalogue is $cz_{\mathrm{obs}} = 3016\pm72\,\mathrm{km\,s^{-1}}$. In both cases the redshift has been multiplied by the speed of light to give units of $\mathrm{km\,s^{-1}}$, but we emphasise that these should not be treated as the total velocity and CMB-frame velocity (because $v_{t} \neq cz_{\mathrm{obs}}$). However, differences in the choice of data and algorithm used to construct the group catalogue can lead to differences in the group observed redshift larger than the uncertainty quoted above. Table~\ref{tab:recvel} presents a number of estimates for the group observed redshift, with and without conversion to the CMB frame, of NGC4993. There is significant overlap between the data used to construct these groups, and many of these rely on data from the 2MASS Redshift \citep{Huchra2012}, 6-degree Field Galaxy Redshift \citep{Jones2009}, and Sloan Digital Sky \citep{York2000} surveys. However, they differ in the clustering algorithms used, their treatment of the various selection effects in the data including if any fainter objects are included, and how the group properties including mean velocities and velocity dispersions are calculated. This is reflected in the variation in the quantities presented in Table~\ref{tab:recvel} even for group catalogues that are built from similar data.

Some of the estimates of the observed redshift of the group containing NGC4993 in Table~\ref{tab:recvel} can be identified as spurious. For instance the estimate using the \cite{Crook2008} Low Density Catalogue (LDC) gives a value significantly lower than other sources. As pointed out in \cite{Hjorth2017}, inspection of the galaxies in the \cite{Crook2008} LDC group shows that NGC4993 is only peripherally associated with this structure, as can also be inferred from the large distance between NGC4993 and the group centre given in Table~\ref{tab:recvel}. \cite{Hjorth2017} provide their own estimate of the group total velocity using a refined version of the \cite{Kourkchi2017} catalogue and using the velocity dispersion of the group as the error to account for the possibility that the group is unrelaxed (both of which are provided in Table~\ref{tab:recvel}). However, the fact is that even after removing possible outliers and inflating the errors, there exists a wide range of `reasonable' observed redshifts that one could use in estimating $H_{0}$. The choices of data or outliers made when computing this adds uncertainty beyond the errors we would naively assume when inferring $H_{0}$. Given the difficulty in deciding which of the values is `correct', this uncertainty should be folded into the analysis. 

\section{Estimates of the peculiar velocity of the group containing NGC4993}\label{sec:pecvel}
In this section we turn to the arguably even less clear case of measuring the peculiar velocity (PV) of the group containing NGC4993. The peculiar velocity of a galaxy encodes the motion induced through gravitational attraction towards dense regions of the universe in departure from the Hubble expansion. Typical line-of-sight PVs are several hundred kilometres per second, and can reach over $1000\,\mathrm{km\,s^{-1}}$ for satellite galaxies in dense regions of the universe where non-linear motions are important. Such a velocity would contribute a substantial fraction of the total velocity for a group of galaxies such as the one hosting NGC4993, and so this absolutely must be considered when making inferences from Standard Sirens.

We clarify that here we are dealing with estimates of the peculiar velocity \textit{of the group} containing NGC4993, not of NGC4993 itself. This is because we are using the observed redshift \textit{of the group} which should account for/remove some of the peculiar velocity, and makes the results less susceptible to the effects of non-linear motions and large peculiar velocities. If one had an accurate measurement of the peculiar velocity of the host galaxy itself, you could use the observed redshift and peculiar velocity of the host alone to constrain $H_{0}$. Unfortunately, such an independent PV measurement does not exist for NGC4993, and even if it did the large error would likely render it unusable. Using group properties instead allows us to average over multiple PV measurements, reducing the errors and the impact of spuriously fast-moving galaxies. It does mean however that one should take differences in group and PV catalogues into account, as we do in this work.

\subsection{Peculiar velocity preliminaries}

Direct measurements of the peculiar velocities of galaxies can be made using a number of different methods including the Tully-Fisher relation \citep{Tully1977}, Fundamental Plane \citep{Dressler1987,Djorgovski1987}, Type-Ia supernovae \citep{Phillips1993}, surface brightness fluctuations \citep{Tonry1988} and the tip of the red-giant branch \citep{Lee1993}. However, those that tend to give the most accurate measurements also tend to be the hardest to obtain and least abundant. The largest single source of PVs to date is the 6-degree Field Galaxy Redshift Survey velocity sample (6dFGSv; \citealt{Magoulas2012}), which contains 8,885 Fundamental Plane galaxies, however the typical uncertainties on the peculiar velocities in this catalogue are $\sim 26\%$ and the hemispherical sky coverage leaves the measurements vulnerable to unknown systematics \citep{Qin2018}. The recently completed 2MASS Tully-Fisher survey (2MTF; \citealt{Hong2019}) has slightly better errors and a more homogeneous coverage, but only 2,062 galaxies. The Cosmicflows-III compilation (CF3; \citealt{Tully2016}) containing 17,669 entries, is currently the largest collection of peculiar velocity measurements. Individual measurements in here come from a variety of sources and techniques (including 6dFGSv) and so can be relatively accurate, but again the inhomogeneous selection and patchwork nature of the catalogue increases the potential for systematics. 

In this work we will use all three of these catalogues to estimate the peculiar velocity of NGC4993. 2MTF and 6dFGSv provide measurements of the `log-distance' ratio $\eta=\mathrm{log_{10}}(d(z_{\mathrm{cmb}})/d(\bar{z}))$; the ratio of the distance inferred from the CMB-frame redshift to the true comoving distance, where the latter is inferred from a distance indicator. This quantity is used because it is linearly related to the change in magnitude induced by a peculiar velocity (which is close to the true observed quantities for the Tully-Fisher and Fundamental Plane relationships) and is close to Gaussian distributed. The full conversion from a log-distance ratio, or any of the true observed quantites, to a PV $\langle v^{\mathrm{obj}}_{\mathrm{p}} \rangle$, results in a PDF that is closer to log-normal, with mean and maximum likelihood values that are biased with respect to the true underlying velocity (see \citealt{Scrimgeour2016} for an excellent example of this). To preserve the Gaussianity of the measurements, one could instead use the estimator of \cite{Watkins2015}
\begin{equation}
\langle  v^{\mathrm{obj}}_{\mathrm{p}} \rangle = \frac{cz_{\mathrm{mod}}}{1+z_{\mathrm{mod}}}\mathrm{ln}(10)\eta,
\end{equation}
where
\begin{equation}
z_{\mathrm{mod}} = z_{\mathrm{cmb}}\biggl[1+\frac{1}{2}(1-q_{0})z_{\mathrm{cmb}} - \frac{1}{6}(1-q_{0}-3q^{2}_{0}+j_{0})z^{2}_{\mathrm{cmb}}\biggl].
\label{eq:zmod}
\end{equation}
For our fiducial cosmology the jerk parameter $j_{0}=1$, whilst the deceleration parameter $q_{0}=\frac{1}{2}(\Omega_{m}-2\Omega_{\Lambda})=-0.55$. When quoting peculiar velocities in this work, these are obtained using this estimator.

\begin{figure}
\centering
\includegraphics[width=0.49\textwidth, trim=0pt 15pt 0pt 0pt, clip]{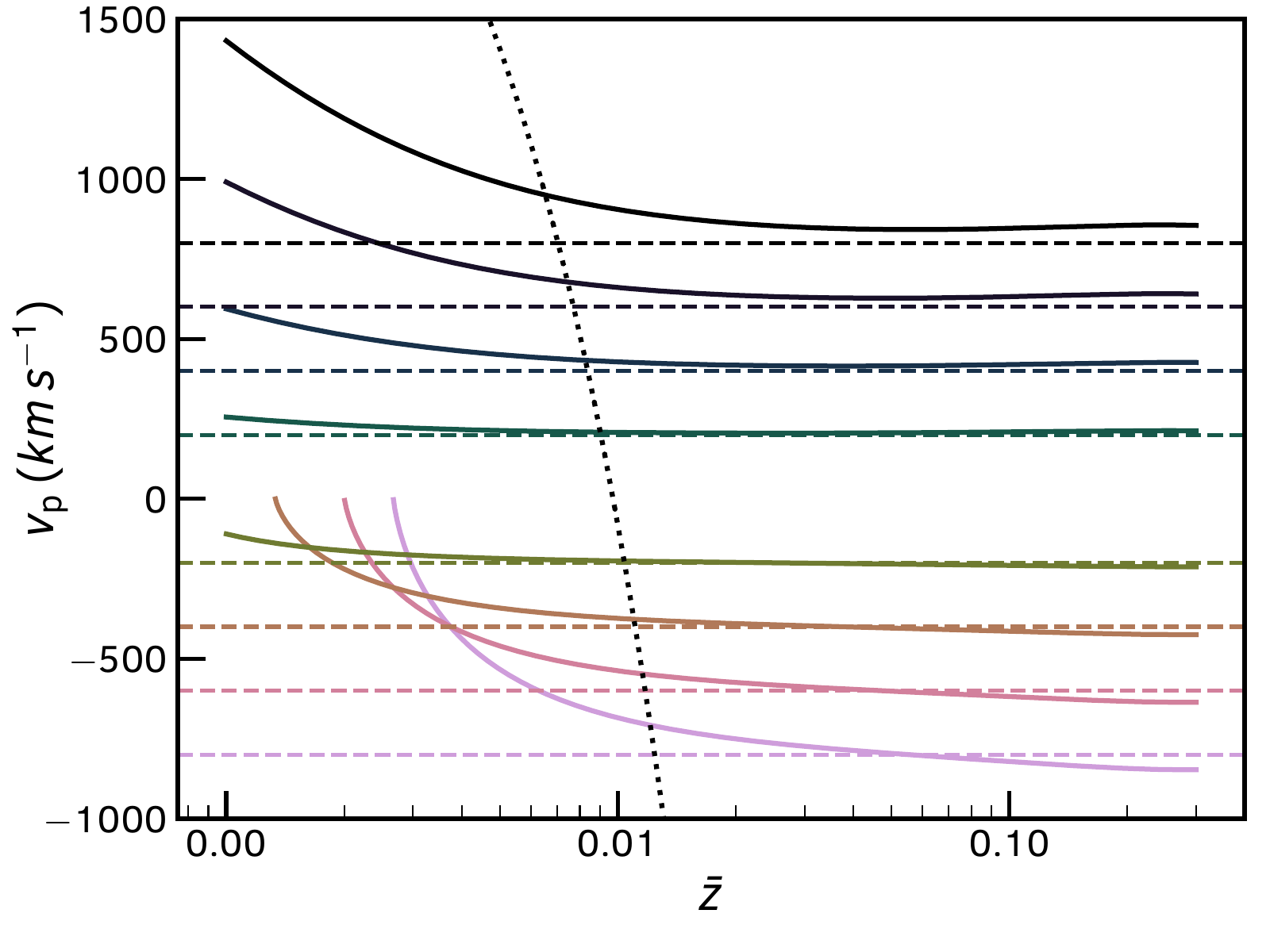}
  \caption{The performance of the \protect\cite{Watkins2015} peculiar velocity estimator as a function of cosmological redshift and true peculiar velocity. The dashed lines show the input velocities, the solid lines are the estimated velocities. The dotted (almost) vertical line denotes a line of constant observed redshift (which is also a constant CMB-frame redshift), in this case that of NGC4993, $z_{\mathrm{obs}}=0.009727$, $z_{\mathrm{cmb}}=0.01077$. The \protect\cite{Watkins2015} estimator does not perfectly recover the true velocity given measurements of the log-distance ratio such as those from modern peculiar velocity catalogues, and is biased at low redshift when $ v^{\mathrm{obj}}_{\mathrm{p}} \ll cz_{\mathrm{cmb}}$ is not satisfied, and at high redshift where it makes the implicit assumption $v_{t}=cz_{\mathrm{obs}}$.}
  \label{fig:pvtest}
\end{figure}

However, this estimator performs poorly under certain conditions and may introduce unwanted systematics into the estimation. This is shown in Figure~\ref{fig:pvtest} where we compute the log-distance ratio, and then the \cite{Watkins2015} velocity, for a range of peculiar velocities and cosmological redshifts. There is a bias at low redshift arising due to the fact that the derivation of the estimator requires $ v^{\mathrm{obj}}_{\mathrm{p}} \ll cz_{\mathrm{cmb}}$ and at high redshift related to the break-down of the assumption that $v_{\mathrm{t}}=cz_{\mathrm{obs}}$ \citep{Davis2019}. For a galaxy at the observed redshift of NGC4993 (denoted by the dotted line in Figure~\ref{fig:pvtest}) the bias is between $5-7\%$ and $10-15\%$ for peculiar velocities of magnitude $400$ and $800\,\mathrm{km\,s^{-1}}$ respectively. These velocities are perfectly reasonable for a galaxy such as NGC4993 and although the bias is small compared to the typical peculiar velocity error this can be avoided entirely by working with the log-distance ratios directly, as we will demonstrate in Section~\ref{sec:hubble}.

Finally, CF3 presents data as distance moduli converted using a consistent cosmological model for all datasets and averaged over the individual measurements if more than one measurement is available for a given galaxy. To convert these to velocities, we first revert the distance moduli to log-distance ratios using the cosmology assumed in CF3 (flat $\Lambda$CDM with $\Omega_{m}=0.27$, $H_{0}=75\,\mathrm{km\,s^{-1}\,Mpc^{-1}}$).

\subsection{Peculiar velocity estimates}

In order to estimate the PV for the group containing NGC4993 and GW170817 we need to average or interpolate over a set of nearby measurements. \cite{Abbott2017c} achieve this by placing a Gaussian weighting kernel of width $8h^{-1}\mathrm{Mpc}$ within the 6dFGSv catalogue, centred on the position of NGC4993. From this they quote a peculiar velocity of $\langle  v^{\mathrm{obj}}_{\mathrm{p}} \rangle = 310 \pm 69\,\mathrm{km\,s^{-1}}$. The uncertainty $\sigma_{v^{\mathrm{obj}}_{\mathrm{p}}}$ is inflated to $150\,\mathrm{km\,s^{-1}}$ to account for potential systematics in their canonical model for extracting $H_{0}$, and this is increased further to $\sigma_{v^{\mathrm{obj}}_{\mathrm{p}}}=250\,\mathrm{km\,s^{-1}}$ in their Appendix to test the robustness of their results.

However, there are a number of choices to be made when estimating the peculiar velocity for NGC4993. Firstly, the Gaussian kernel used in \cite{Abbott2017c} is larger than the typical size of the groups (and the distance of NGC4993 from the center of those groups) identified in Section~\ref{sec:recvel}/Table~\ref{tab:recvel} above and so a smaller scale weighting scheme might be more appropriate. Perhaps the most obvious way of ensuring consistency between the group total and peculiar velocities is instead to average only PV measurements for galaxies within the group. There is also the question of which catalogues to use in the estimation, weighing the large number of measurements in, for instance, CF3 with the clearer selection function of 2MTF. The best solution to this problem is not obvious, and as we will show below, different choices lead to substantial changes in the peculiar velocity which in turn impact the final constraints on $H_{0}$ even for the single event considered here. 

We start by identifying which of the groups in Section~\ref{sec:recvel} contain objects with direct PV measurements. The number of such measurements is given in Table~\ref{tab:recvel} as is the average peculiar velocity. Unfortunately, the number is limited; the only group with a suitably large number is \cite{Crook2008} (LDC), however the velocities for this group may not  be representative of NGC4993 given its distance from the group centre (although this is still smaller than the smoothing scale used in \citealt{Abbott2017c} as mentioned above). Another promising case is that of \cite{Tully2015}, which contains three measurements in the CF3 catalogue. Although the uncertainty is quite large, we identify this combination of group observed redshift and peculiar velocity as one `reasonable' choice.

\begin{figure}
\centering
\includegraphics[width=0.49\textwidth, trim=0pt 40pt 0pt 0pt, clip]{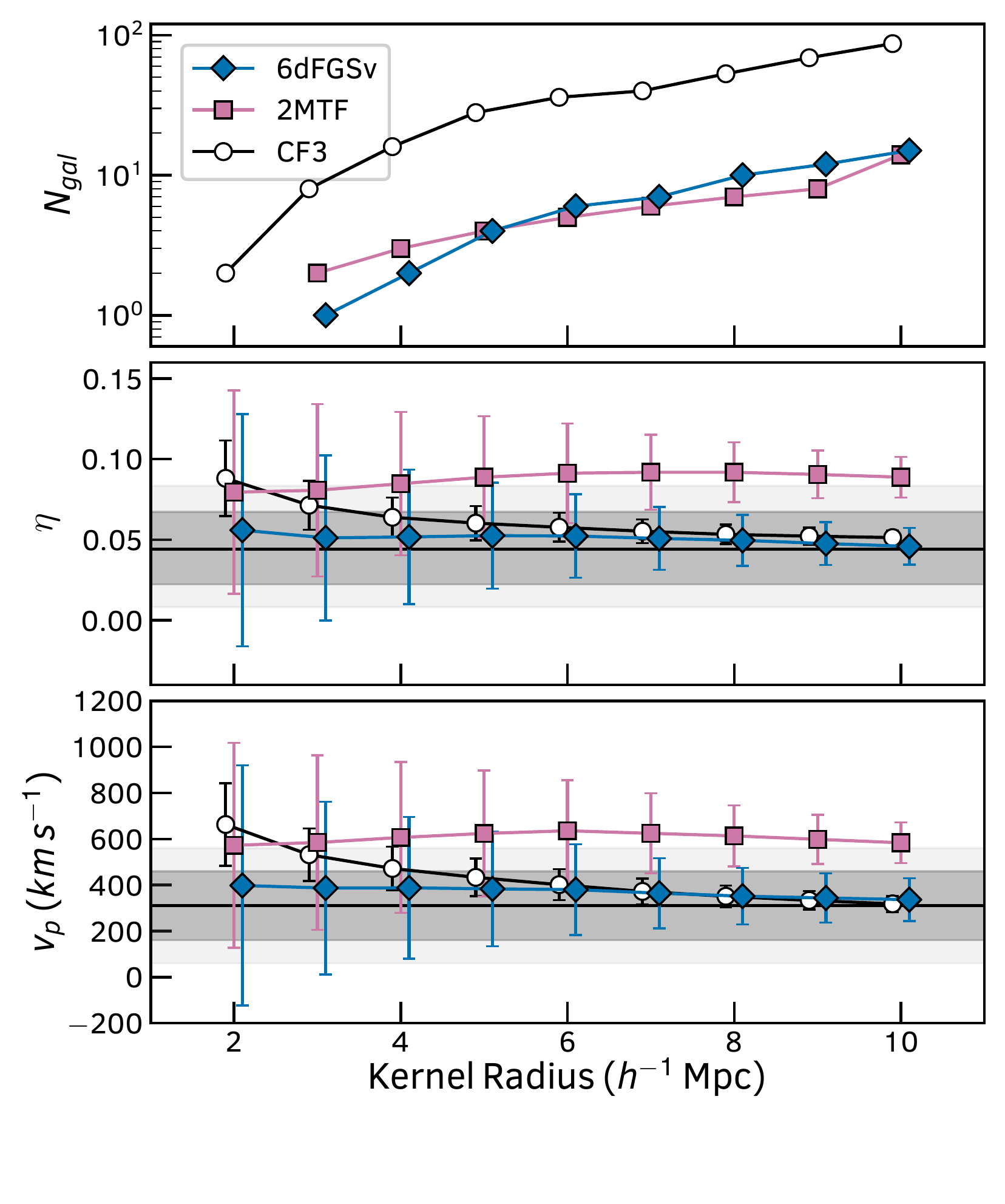}
  \caption{Peculiar velocities at the position of NGC4993 computed using Gaussian kernels of varying radii.  Different points correspond to different catalogues. The top panel shows the number of PV measurement in the catalogue within one kernel radius of NGC4993 . The middle panel shows the weighted mean and statistical uncertainty on the log-distance ratios. The bottom panel shows the weighted mean and statistical uncertainty on the PV from each catalogue calculated from the middle panel using the estimator of \protect\cite{Watkins2015}. The darker and lighter solid bands show the value used in \protect\cite{Abbott2017c} with errors of $150$ and $250\,\mathrm{km\,s^{-1}}$ respectively. In the middle panel we have propagated the velocity distribution used in \protect\cite{Abbott2017c} to that of log-distance ratios and used the $16^{\mathrm{th}}, 50^{\mathrm{th}}$ and $84^{\mathrm{th}}$ percentiles, which is necessary as the transformation in non-Gaussian.}
  \label{fig:pecvel}
\end{figure}

We next turn to alternative smoothed estimates of the velocity field, using the CF3, 2MTF, and 6dFGSv catalogues. In Fig.~\ref{fig:pecvel} we show the log-distance ratio and peculiar velocity at the position of NGC4993 computed using a Gaussian kernel with width varying between $2-10 h^{-1}\,\mathrm{Mpc}$. On the same figure we plot the peculiar velocity and error budget used by \cite{Abbott2017c} and the corresponding range of $\eta$ values. We find differences between measurements obtained used different catalogues that are larger than the error bars on these measurements, or those adopted by \cite{Abbott2017c}, would suggest.

The obvious difference in Figure~\ref{fig:pecvel} is between the 2MTF and CF3/6dFGSv catalogues. We identified that this is due to a small global offset in the distances estimated from both catalogues. \cite{Qin2019} compare the 1,096 common galaxies between 2MTF and CF3 (the CF3 measurements typically come from older measurements which have been updated in, and superseded by, 2MTF). They find a relationship of $\mathrm{log_{10}}(d_{\mathrm{CF3}}(\bar{z})) = 0.96\mathrm{log_{10}}(d_{\mathrm{2MTF}}(\bar{z})) + 0.08$ where $d(\bar{z})$ is the comoving distance to the galaxy computed from the distance indicators in these catalogues. Correcting the 2MTF log-distance ratios using this equation and our fiducial cosmology, we recover PVs for NGC4993 from 2MTF that are a much closer match to the CF3 estimate across all Gaussian kernels. This does not provide a solution to our problem however. The source of this discrepancy is unclear; we would expect the 2MTF measurements for each galaxy to be more up-to-date and robust compared to those for the same object in CF3 and there are arguments for and against the zero-point calibration in both catalogues. 2MTF has a much more homogeneous distribution and selection of galaxies compared to CF3 but a smaller depth, which would both decrease and increase the relative zero-point uncertainty respectively. Again, given the difficulty in choosing which of these PV measurements is better, and the mere fact that a small calibration offset can create such a large difference, we should fold the uncertainty into our Standard Siren measurements.

\section{Changes in the Hubble constant} \label{sec:hubble}
In this section we reproduce the analysis of \cite{Abbott2017c} with different measurements of the total and peculiar velocities for GW170817 to investigate the effects on the $H_{0}$ constraints. Our method of computing the posterior probability of $H_{0}$ given the observed gravitational-wave event and measured velocities after marginalising over all relevant quantities is modified compared to that used in their work. The method we use here has two benefits. Firstly, it uses the actual observed redshift $z_{\mathrm{obs}}$ and the full, correct relationship between the various redshifts given in our Eq.~\ref{eq:redshiftproper} rather than the approximation used in \cite{Abbott2017c} (our  Eq.~\ref{eq:redshiftapprox}). Given the low redshift of NGC4993 we expect this to have a small effect on the recovered $H_{0}$ constraints, but avoiding such approximations is preferred and does not increase the complexity of the model. 

Secondly, we do not use peculiar velocities in our model and stick instead with log-distance ratios, which are much closer to the observed quantities from peculiar velocity surveys and have statistical uncertainties that are closer to being naturally Gaussian distributed. Again, this does not significantly increase the model complexity and we argue is a more natural choice for Standard Siren measurements. Perhaps the only caveat to this is if the estimates of the peculiar velocity come from reconstructed fields, i.e., that of \cite{Carrick2015, Graziani2019}, where it is typically assumed the velocity field is Gaussian. In this case a posterior based on the model velocity (i.e., similar to that used by \citealt{Abbott2017c}) may be more appropriate. See the Appendix for our version of the posterior in this scenario. It is worth noting however that such reconstructions are based either on measurements of the density field (which is non-Gaussian), or the same peculiar velocity surveys we are using directly (which also result in non-Gaussian velocities). Hence the assumption of a Gaussian velocity field from these reconstructions could introduce systematic errors.

We start by writing the posterior probability
\begin{align}
&p(H_{0}|x_{GW},\langle \eta \rangle,\langle z_{\mathrm{obs}} \rangle,\langle z^{\mathrm{Sun}}_{\mathrm{p}} \rangle) \propto p(H_{0}) \int \bigg[ p(x_{GW}|d_{L},\mathrm{cos}\,\iota) \notag \\
&\times p(\langle \eta \rangle|H_{0},d_{L},z_{\mathrm{obs}},z^{\mathrm{Sun}}_{\mathrm{p}})\,p(\langle z_{\mathrm{obs}} \rangle|z_{\mathrm{obs}})\,p(\langle z^{\mathrm{Sun}}_{\mathrm{p}} \rangle|z^{\mathrm{Sun}}_{\mathrm{p}}) \notag \\
& \,\,\,\,\times p(d_{L})\,p(\mathrm{cos}\,\iota)\,p(z_{\mathrm{obs}})\,p(z^{\mathrm{Sun}}_{\mathrm{p}})\bigg]\,\mathrm{d}d_{L}\,\mathrm{dcos}\,\iota\,\mathrm{d}z_{\mathrm{obs}}\,\mathrm{d}z^{\mathrm{Sun}}_{\mathrm{p}},
\label{eq:h0posterior}
\end{align} 
where $d_{L}$ and $\mathrm{cos}\,\iota$ are the luminosity distance and inclination of the GW event inferred from the event itself, $\langle z_{\mathrm{obs}} \rangle$, $\langle z^{\mathrm{Sun}}_{\mathrm{p}} \rangle$ and $\langle \eta \rangle$ are the measured observed redshift of the host galaxy (or the group it is in), the Sun's peculiar redshift corresponding to the CMB dipole velocity in the direction of the host (or group) and the measured log-distance ratio respectively. $x_{GW}$ are the observations of the gravitational waves made by the LIGO \citep{LIGO2015} and Virgo \citep{Acernese2015} detectors. Hence, $p(x_{GW}|d,\mathrm{cos}\,\iota)$ denotes the likelihood of the observations given the distance and inclination of the merger (all other parameters related to the gravitational waveform observations, for instance the masses of the two inspiralling sources, have already been marginalised out). The uncertainty on the velocity and direction of the CMB dipole is extremely small \citep{Planck2018} and so we treat $p(\langle z^{\mathrm{Sun}}_{\mathrm{p}} \rangle|z^{\mathrm{Sun}}_{\mathrm{p}})$ as a $\delta$ function.\footnote{We have included it in our posterior calculation as it should be included in the case where the host galaxy is not known. Additionally, we have not considered uncertainty in the angular position of the host group. If we did there would be some uncertainty on $z^{\mathrm{Sun}}_{\mathrm{p}}$ even if the direction and velocity of the CMB dipole were perfectly known.} For the remaining likelihoods, we assume Gaussian distributions, such that
\begin{align}
p(\langle z_{\mathrm{obs}} \rangle|z_{\mathrm{obs}}) &= N[z_{\mathrm{obs}},\sigma_{z_{\mathrm{obs}}}](\langle z_{\mathrm{obs}} \rangle), \\
p(\langle \eta \rangle|H_{0},d_{L},z_{\mathrm{obs}},z^{\mathrm{Sun}}_{\mathrm{p}}) &= N[\eta(H_{0},d_{L},z_{\mathrm{obs}},z^{\mathrm{Sun}}_{\mathrm{p}}),\sigma_{\eta}](\langle \eta \rangle),
\label{eq:recvelpdf}
\end{align}
are the likelihoods for the observed redshift and log-distance ratio. $\sigma_{z_{\mathrm{obs}}}$ and $\sigma_{\eta}$ are the measurement errors of these quantities. The final piece we need is the set of equations to compute the model log-distance ratio given the observed parameters $z_{\mathrm{obs}}$, $z^{\mathrm{Sun}}_{\mathrm{p}}$ and $d_{L}$, and the parameter we want to measure, $H_{0}$. These can be derived from the definition of the log-distance ratio,
\begin{equation}
\eta = \mathrm{log_{10}}\biggl(\frac{d(z_{\mathrm{cmb}})}{d(\bar{z})}\biggl) = \mathrm{log_{10}}\biggl((1+z_{\mathrm{obs}})\frac{d(z_{\mathrm{cmb}})}{d_{L}}\biggl),
\label{eq:modeleta}
\end{equation}
where 
\begin{equation}
d(z_{\mathrm{cmb}}) = \frac{c}{H_{0}}\int_{0}^{\frac{1+z_{\mathrm{obs}}}{1+z^{\mathrm{Sun}}_{\mathrm{p}}}-1}\frac{dz}{E(z)},
\end{equation}
and $E(z) = \sqrt{\sum_i \Omega_i (1+z)^{-3(1+w_i)}}$ is the usual  redshift dependent part of the expansion rate. Herein we assume a Flat $\Lambda$CDM cosmological model, although this could easily be expanded to incorporate alternative cosmological models. The above framework can also be modified to work for model peculiar velocities rather than log-distance ratios, but without the need to make any of the approximations used in \cite{Abbott2017c}. This is shown in the Appendix.

Although the above method may seem more complex than that used in \cite{Abbott2017c}, in practice it does not take much longer to compute (the redshift-distance relationship can be spline interpolated for fast inversion for any value of $H_{0}$). Crucially however, it makes no assumptions about the relationship between the redshifts and velocities of interest, which may otherwise introduce systematic errors, and it works more closely with the observed quantities of the host, namely the group observed redshift and the log-distance ratio. We investigate how our modified posterior compares to that of \cite{Abbott2017c} below. We found no dependence of the results in this work on the assumed value of $\Omega_{m}$, as would be expected given the low redshift of NGC4993.\footnote{For future Standard Sirens at higher redshift where this effect may not be negligible, one could follow the method used in the cosmological analysis of type-Ia supernovae and reduce the cosmological dependence of the fit for the Hubble constant without resorting to the approximation $d(z_{\mathrm{cmb}}) \approx \frac{cz_{\mathrm{cmb}}}{H_{0}}$ by instead using  $d(z_{\mathrm{cmb}}) \approx \frac{cz_{\mathrm{mod}}}{H_{0}}$ with $z_{\mathrm{mod}}$ given by Eq.~\ref{eq:zmod}.}

Rather than repeat the full analysis of the gravitational-wave event to evaluate $p(x_{GW}|d_{L},\mathrm{cos}\,\iota)$ and $p(H_{0}|x_{GW},\langle z_{\mathrm{obs}} \rangle,\langle z^{\mathrm{Sun}}_{\mathrm{p}} \rangle,\langle \eta \rangle)$, we will make use of the posterior samples for GW170817 provided by both the LIGO collaboration\footnote{These can be found at \url{https://dcc.ligo.org/LIGO-P1800061/public/}. We use the samples assuming a `high-spin' prior.} and from \cite{Hotokezaka2019}. The latter includes additional information on the source inclination that significantly strengthens the constraints on $H_{0}$. In using these samples to evaluate Eq.~\ref{eq:h0posterior} we perform an MCMC\footnote{We use the publicly available \textsc{emcee} python routine \citep{Foreman-Mackey2012}.} over the parameters $z_{\mathrm{obs}}$ and $H_{0}$ and at each likelihood evaluation we draw from the marginalised posterior samples for $d_{L}$ and $\mathrm{cos}\,\iota$. The combined posterior from this procedure then approximates the full posterior for all four parameters as if we had fit the GW signal directly. 

\begin{table*}[h!]
\setlength{\extrarowheight}{3pt}
\caption{Constraints on $H_{0}$ from GW170817 and NGC4993 assuming different combinations of observed redshift, and log-distance ratio. Where only a group reference is provided as a description this indicates use of the group observed redshift and mean CF3 log-distance ratio from Table~\ref{tab:recvel}. Where a group reference and a PV catalogue are provided, this indicates the use of the group observed redshift and log-distance ratio from the catalogue using a Gaussian kernel of radius R. For each case we provide the values and errors used in constraining $H_{0}$. The last two columns give the maximum \textit{a posteriori} value and $68\%$ equal likelihood bounds on $H_{0}$ without and with additional constraints on the source inclination angle respectively (and so the left and right sub-columns correspond to the analyses of \protect\cite{Abbott2017c} and \protect\cite{Hotokezaka2019} respectively). The first row is where we use the same group and peculiar velocity catalogues/methods used in these previous works, but with our updated Bayesian model.}
\centering
\begin{tabular}{lccccc} \hline
\multirow{3}{*}{Case No.} & \multirow{3}{*}{Description} & \multirowcell{3}{Observed Redshift \vspace{4pt} \\ $\langle z_{\mathrm{obs}} \rangle$ ($\times 10^{-5}$)} &\multirowcell{3}{Log-distance Ratio \vspace{4pt} \\ $\langle \eta \rangle$} & \multicolumn{2}{c}{Hubble constant} \\ & & &  & \multicolumn{2}{c}{$H_{0}$ ($\mathrm{km\,s^{-1}\,Mpc^{-1}}$)} \\
& & &  & GW & GW+VLBI \\ \hline
1. & Canonical: \cite{Crook2008} (HDC) + 6dFGSv ($R=8h^{-1}\,\mathrm{Mpc}$) & $1006  \pm 24$ & $0.049 \pm 0.023$ & $69.5^{+11.7}_{-7.3}$ &  $67.9^{+4.6}_{-4.5}$ \\
2. & \cite{Tully2015} & $1000 \pm 17$ & $0.114 \pm 0.039$ & $60.1^{+11.3}_{-7.7}$ &  $58.0^{+6.1}_{-5.3}$  \\
3. & \cite{Kourkchi2017} & $999 \pm 11$ & $0.068 \pm 0.051$ & $67.2^{+14.1}_{-10.1}$ &  $64.6^{+8.7}_{-7.5}$ \\
4. & 6dFGRS + CF3 ($R=2h^{-1}\,\mathrm{Mpc}$) & $1010 \pm 17$ & $0.089 \pm 0.024$ & $63.5^{+10.6}_{-6.5}$ &  $62.3^{+4.1}_{-4.1}$ \\
5. & \cite{Kourkchi2017} (Trimmed) + 6dFGSv ($R=8h^{-1}\,\mathrm{Mpc}$) & $974 \pm 17$ & $0.049 \pm 0.023$ & $67.4^{+11.0}_{-7.1}$ &  $65.9^{+4.3}_{-4.3}$ \\
6. & \cite{Crook2008} (HDC) + 2MTF ($R=8h^{-1}\,\mathrm{Mpc}$) & $1006 \pm 24$ & $0.091 \pm 0.023$ &  $63.2^{+10.2}_{-6.7}$ &  $61.7^{+4.1}_{-4.1}$ \vspace{3pt} \\ \hline
\end{tabular}
\label{tab:results}
\end{table*}

The accuracy of this approximation depends on how well the finite number of posterior samples for $d_{L}$ and $\mathrm{cos}\,\iota$ represents the true posterior, which we test by repeating the canonical analyses of \cite{Abbott2017c} and \cite{Hotokezaka2019} (which uses the approximate Eq.~\ref{eq:redshiftapprox}). Doing so, setting $cz_{\mathrm{cmb}}=3327\pm 72\,\mathrm{km\,s^{-1}}$ and $\langle v^{\mathrm{obj}}_{\mathrm{p}} \rangle=310\pm 150\,\mathrm{km\,s^{-1}}$, we find $H_{0}=69.8^{+11.0}_{-7.2}\,\mathrm{km\,s^{-1}\,Mpc^{-1}}$ and $H_{0}=68.3^{+4.5}_{-4.4}\,\mathrm{km\,s^{-1}\,Mpc^{-1}}$ respectively. Our errors are slightly smaller than those found in \cite{Abbott2017c} ($H_{0}=70.0^{+12.0}_{-8.0}\,\mathrm{km\,s^{-1}\,Mpc^{-1}}$) but fully consistent with \cite{Hotokezaka2019} ($H_{0}=68.1^{+4.5}_{-4.3}\,\mathrm{km\,s^{-1}\,Mpc^{-1}}$). We attribute the differences to the finite number of samples in the GW posterior chains. Our estimate of the distance to GW170817 from \citealt{Abbott2017c} is $d_{L} = 44.2^{+2.3}_{-6.5}\mathrm{Mpc}$ (to be compared to the quoted value of $d_{L} = 43.8^{+2.9}_{-6.9}\mathrm{Mpc}$); the small shift in the peak and slight underestimation of the errors is likely because the posterior samples do not fully represent the long tail of the true luminosity distance distribution. This in turn leads to small differences in the recovered $H_{0}$ posterior. For fairness, we compare all our results for different cases below to \textit{our} results for the canonical model, rather than the quoted results. 

\subsection{Comparing Bayesian models}

\begin{figure*}[h!]
\centering
\subfloat{\includegraphics[width=0.49\textwidth]{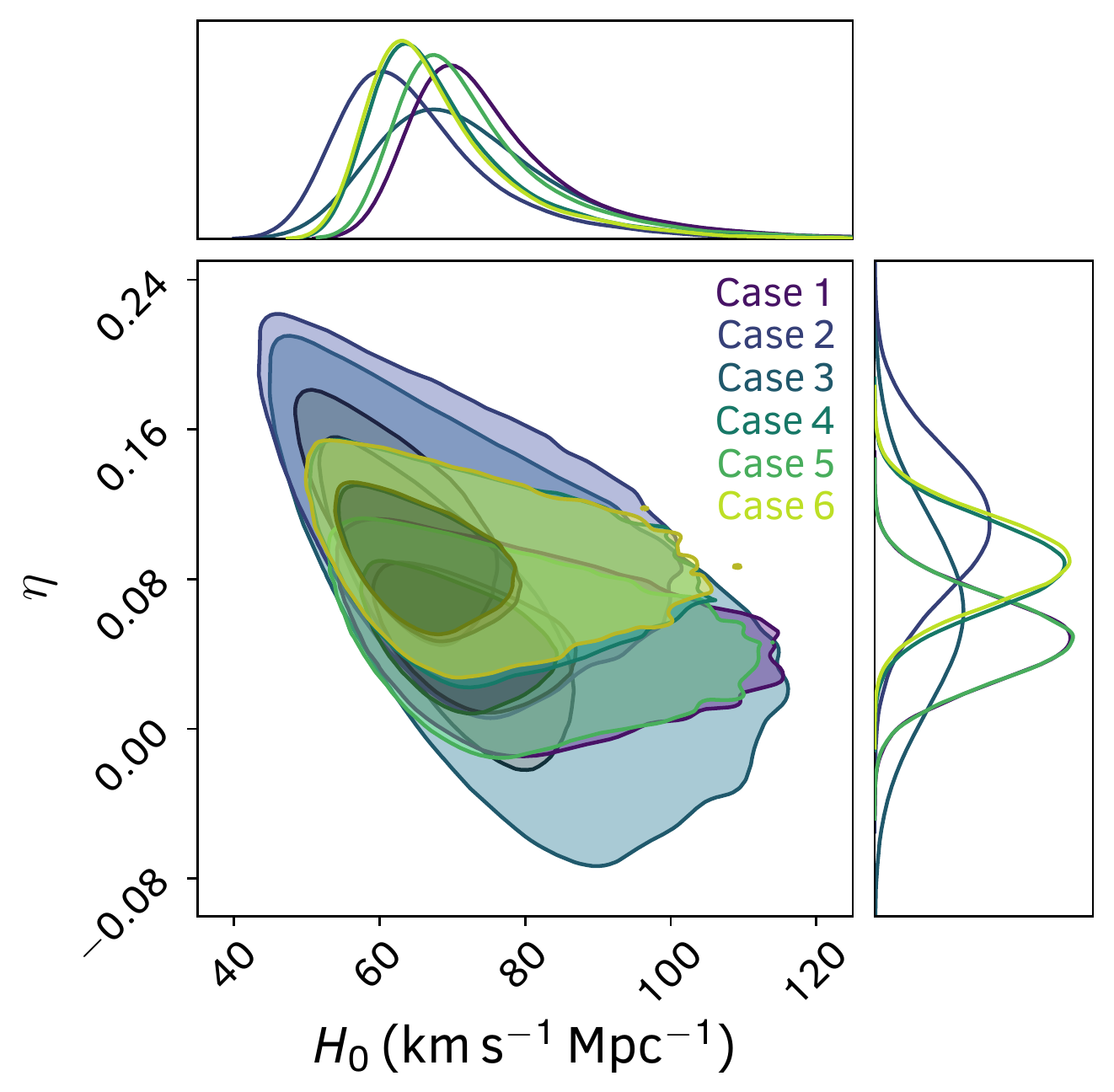}}
\subfloat{\includegraphics[width=0.49\textwidth]{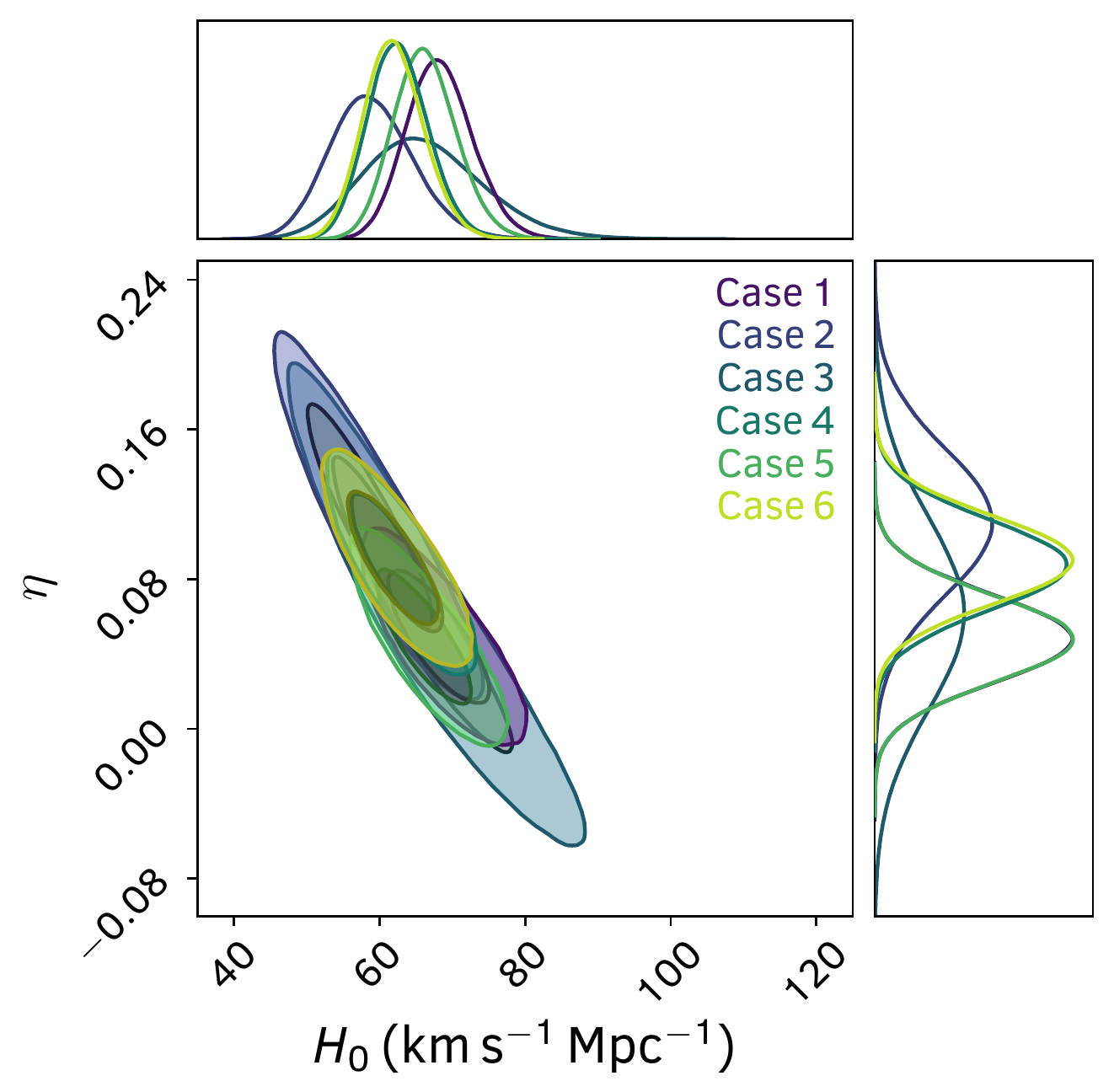}}
  \caption{Marginalised constraints on the Hubble constant $H_{0}$, and corresponding model log-distance ratios $\eta$, of NGC4993 from the Standard Siren measurement of GW170817 for the 6 different cases listed in Table~\ref{tab:results}. The left plot shows the constraints without any additional information on the viewing angle \protect\citep{Abbott2017c}, whilst right shows the tighter constraints when this information is included \protect\citep{Hotokezaka2019}.}
  \label{fig:results}
\end{figure*}

We begin by comparing the results using our preferred Bayesian model presented above, to that used in \cite{Abbott2017c} and \cite{Hotokezaka2019}. In this fit (and all fits from here on) we use a prior $p(H_{0}) \propto 1/H_{0}$, a volumetric prior $p(d_{L}) \propto d^{2}_{L}$ and a flat prior $p(z_{\mathrm{obs}})$ with $cz_{\mathrm{obs}} \in [2000, 4000]\,\mathrm{km\,s^{-1}}$. For comparison we start with $\langle cz_{\mathrm{obs}} \rangle = 3016\pm72$ and $\langle \eta \rangle = 0.049 \pm 0.023$, which are the values derived using exactly the same group and peculiar velocity catalogue as \cite{Abbott2017c} and where we have adopted an error on the log-distance ratio that is close to that used for the velocity in \cite{Abbott2017c} (applying the \cite{Watkins2015} estimator to this value for the error returns $154\,\mathrm{km\,s^{-1}}$).

Using the above procedure and our fiducial cosmology, we find $H_{0}=67.9^{+4.6}_{-4.5}\,\mathrm{km\,s^{-1}\,Mpc^{-1}}$and $H_{0}=69.5^{+11.7}_{-7.3}\,\mathrm{km\,s^{-1}\,Mpc^{-1}}$ with and without the extra constraints on the source inclination respectively. There is a small decrease in the maximum \textit{a posteriori} values of $<0.5\,\mathrm{km\,s^{-1}\,Mpc^{-1}}$ compared to the canonical model, which is not significant compared to the uncertainties and the $68\%$ equal likelihood bounds cover a nearly identical range of $H_{0}$ values. However, we emphasise again that our model makes fewer assumptions and is closer to the observed quantities and so will be preferable as more gravitational waves are detected, constraints on the Hubble constant become tighter and the impact of potential systematic errors becomes more important.

\subsection{Different choices of recession and peculiar velocity}

We next look at the results of using alternative estimates of the total and peculiar velocities of NGC4993 when constraining $H_{0}$. We test a wide range of cases based on the results of Sections~\ref{sec:recvel} and~\ref{sec:pecvel}, firstly using the observed redshift and mean log-distance ratio for seven of the groups listed in Table~\ref{tab:recvel} (all but \citealt{Crook2008} (LDC), which as explained in Section~\ref{sec:recvel} has a redshift far from that of NGC4993), and then combining these different group observed redshifts with the log-distance ratios calculated from the three peculiar velocity catalogues in Section~\ref{sec:pecvel} with Gaussian Kernels of different widths. We ultimately obtained posterior distributions for the Hubble constant for all possible combinations of group catalogue, peculiar velocity survey and Gaussian Kernel with width between $2-8 h^{-1}\,\mathrm{Mpc}$, which combined with the constraints using only the group catalogues gives a total of 154 different scenarios. In all cases we place a minimum possible error on the observed redshift (times the speed of light) of $50\,\mathrm{km\,s^{-1}}$ and on the log-distance ratio of $0.023$, similar to what was done in \cite{Abbott2017c} and mimicking one potential way that we might try (unsuccessfully, as we will show) to mitigate potential systematics. 

Descriptions of 5 of the additional cases, any of which we argue could equally be chosen when analysing event GW170817,  are presented in Table~\ref{tab:results} along with the resulting $H_{0}$ constraints with and without the extra information on the source inclination from \cite{Hotokezaka2019}. In Figure~\ref{fig:results}, we show the marginalised posteriors on $H_{0}$ and the model log-distance ratio of NGC4993 for these cases. We note that the value of $\eta$ given in these figures are those computed for each of the posterior samples using Eq~\ref{eq:modeleta}. By definition the distribution of these should match the assumed probability distribution of the observed log-distance ratio. We include these in Figure~\ref{fig:results} to highlight the strong dependence of the Hubble constant on the choice of observations.

From the results presented in this Figure we see that different choices or determinations of the total and peculiar velocities of NGC4993 result in different constraints that, while consistent at the $68\%$ confidence level (although only just for the more constrained data from \citealt{Hotokezaka2019}), have a spread larger than the error from the Canonical analysis (Case 1) would suggest. It is also interesting to note that of the possible cases presented here (and compared to the majority of the alternative cases we tested) the canonical analysis produces the largest $H_{0}$ constraints. This is a result of the relatively large observed redshift and small log-distance ratio (peculiar velocity) used in this analysis. 

\begin{figure*}
\centering
\subfloat{\includegraphics[width=0.49\textwidth]{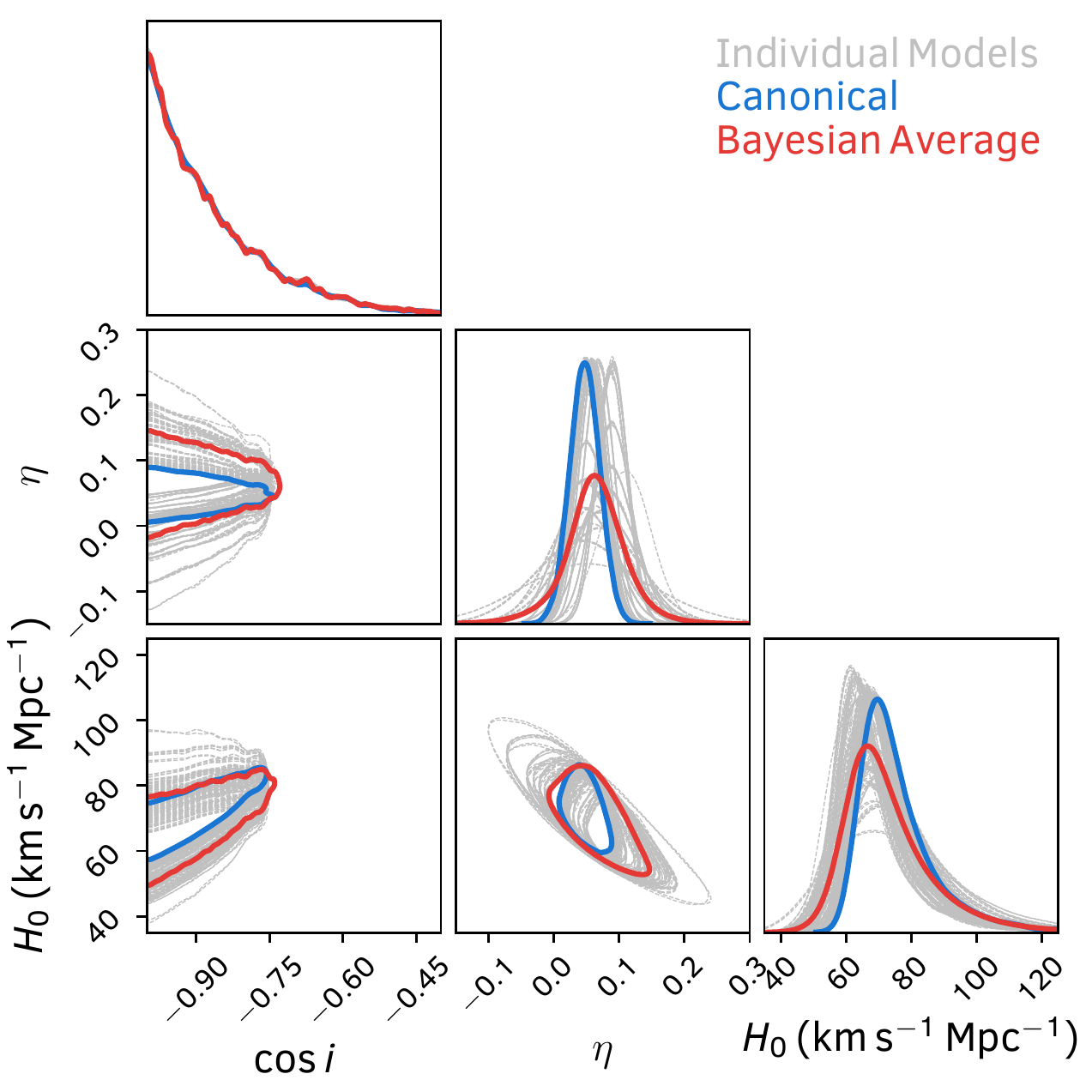}}
\subfloat{\includegraphics[width=0.49\textwidth]{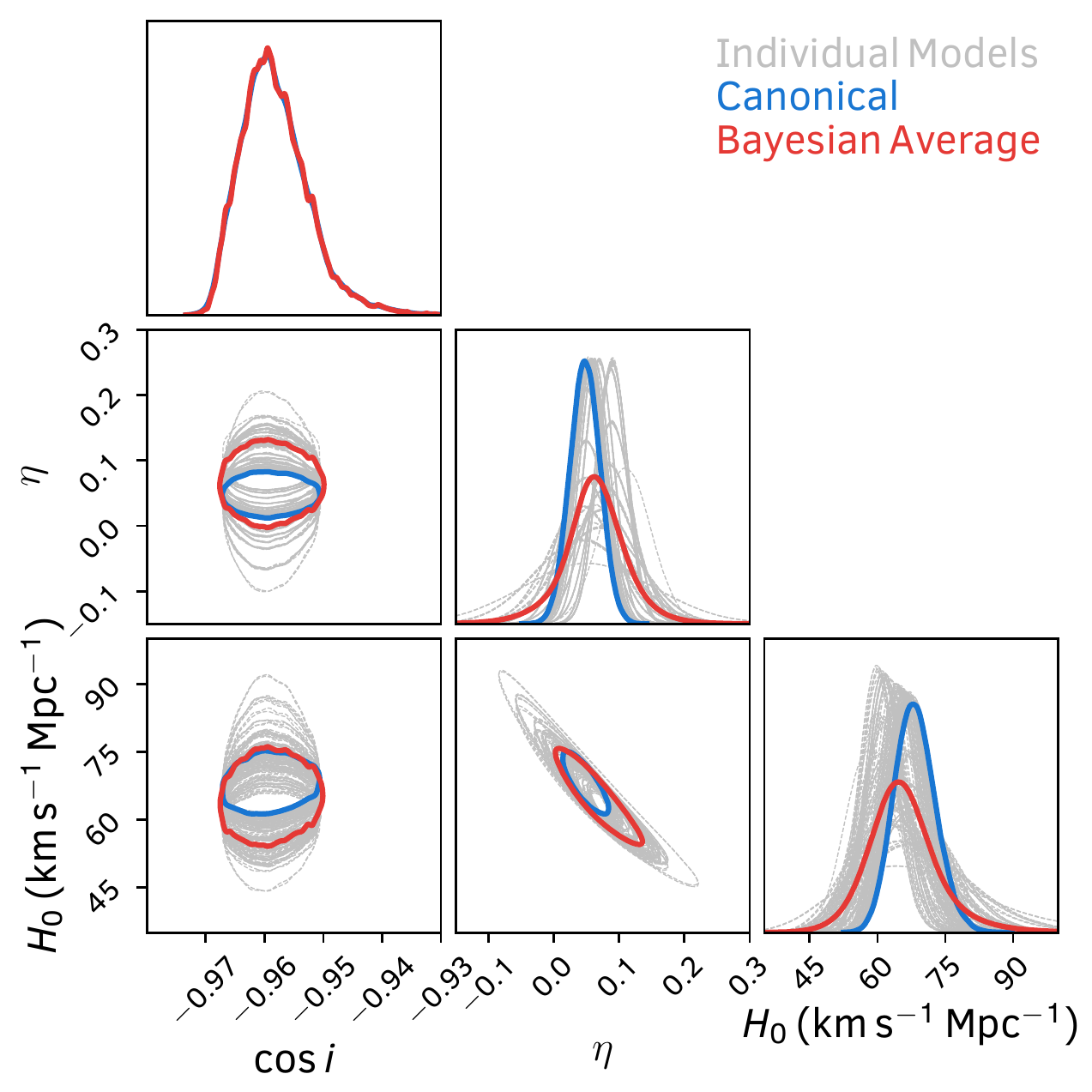}}
  \caption{Constraints on the Hubble constant $H_{0}$ and GW source inclination $\mathrm{cos}\,\iota$, and the corresponding model log-distance ratio of NGC4993 $\eta$ using Bayesian Model Averaging to encapsulate our uncertainty on the different measurements of the total observed velocity and log-distance ratio. The left plot shows the constraints without any additional information on the viewing angle (i.e., corresponding to \protect\citealt{Abbott2017c}), whilst right shows the tighter constraints when this information is included \protect\citep{Hotokezaka2019}. In both cases, the canonical model from these references is shown in blue, the various individual models from this work (using different combination of observed redshift and log-distance ratio) are in gray. The Bayesian Average of the individual models is shown in red. Performing a weighted average of the different models more accurately accounts for our uncertainty in the observed quantities, giving more robust constraints.}
  \label{fig:combined}
\end{figure*}

It is clear that in order for robust constraints to be obtained we need to better understand which of the techniques for calculating the group observed redshift and log-distance ratio or peculiar velocity gives the most reliable and accurate results, or better account for our uncertainty on these quantities. This is more complicated than simply increasing the standard deviation on our measurement of the peculiar velocity. For the case of GW170817 alone this is demonstrated by the fact that we have included conservative lower limits on the errors for all our different combinations and that \cite{Abbott2017c} inflated the peculiar velocity errors even further to test the robustness of their constraints, but these results still do not encompass the spread seen in Fig.~\ref{fig:results}. In the following section, we instead suggest the use of Bayesian Model Averaging (see \citealt{Parkinson2013} for a review) to combine all the cases we test here into a more representative constraint on $H_{0}$.

\subsection{Bayesian Model Averaged Hubble constant}
One way to evaluate the relative weight we should give to each $H_{0}$ posterior when dealing with uncertainty in the true observed redshift or log-distance ratio of the source is the Bayesian Evidence. For a given model $\mathcal{M}_{i}$, this is the probability of the model given our data $p(\mathcal{M}_{i}|D)$. In our scenario we have 154 different models corresponding to our seven different group catalogues, three peculiar velocity surveys and seven kernel radii. Bayesian Model Averaging uses the Evidence as a way to perform a weighted average over all the models and create a unified posterior $p(\theta|D)$ that accounts for our uncertainty regarding which model is correct. Mathematically, we write this in terms of the parameter posteriors $p(\theta|D,\mathcal{M}_{i})$ for each model and the Evidence, 
\begin{equation}
p(\theta |D) = \frac{\sum_{i}p(\theta | D,\mathcal{M}_{i})p(\mathcal{M}_{i}|D)}{\sum_{i}p(\mathcal{M}_{i}|D)}.
\end{equation}

We first compute the Evidence for each of the cases.\footnote{We do this using the implementation of Single Ellipsoid Nested Sampling \citep{Mukherjee2006} in the python package \textsc{nestle} found here: \url{http://kylebarbary.com/nestle/}. The way nested sampling works means we can no longer sample from the posterior chains for GW170817 to obtain $\mathrm{cos}\,\iota$ and $d_{L}$ inside the likelihood evaluation. Instead, we use Gaussian Kernel Density Estimation (KDE) to reproduce the 2D likelihood surface of these two parameters and perform the nested sampling over these, $H_{0}$ and $z_{\mathrm{obs}}$ (so 4 dimensions in total), evaluating the likelihood using the KDE for $\mathrm{cos}\,\iota$ and $d_{L}$ at every point. This produces results consistent with the sampling method used previously. Future studies could incorporate the Evidence calculation directly when fitting the GW data, but the approximate method here is suitable for the illustrative purposes of this work.} We then produce a combined posterior by weighting the samples from each individual model by the Evidence. The result of this procedure is shown in Fig.~\ref{fig:combined}. The combined posterior more fully represents our true uncertainty on the Hubble constant. We obtain averaged marginalised constraints of $H_{0}=66.8^{+13.4}_{-9.2}\,\mathrm{km\,s^{-1}\,Mpc^{-1}}$; and $H_{0}=64.8^{+7.3}_{-7.2}\,\mathrm{km\,s^{-1}\,Mpc^{-1}}$ when additional information on the viewing angle is included. Whilst these are consistent with the canonical results of \cite{Abbott2017c} and \cite{Hotokezaka2019}, accounting for our uncertainty in the total observed and peculiar velocities of NGC4993 has decreased the maximum \textit{a posteriori} value in both cases by $\sim 3\,\mathrm{km\,s^{-1}\,Mpc^{-1}}$ because the majority of models have lower observed redshifts and larger log-distance ratios than the canonical model. 

Averaging over our model uncertainty has also increased the width of the $68\%$ confidence interval by $\sim 35\%$ and $\sim68\%$ without and with the extra information on the source inclination respectively. In the latter case, with the source inclination known much more precisely, the unknown peculiar velocity of NGC4993 is the dominant source of uncertainty and has the potential to strongly bias our results. Performing an average over all the various measurements should mitigate against this quite well, but substantially increases the uncertainty on the Hubble constant. This demonstrates that the accuracy with which we can measure and model the velocity field in the local Universe will remain an important consideration in the error budget for future Standard Siren measurements and is an area where improvement could be made. From Figure~\ref{fig:combined} it is also apparent that simply increasing the uncertainty on the log-distance ratio or peculiar velocity, as tested in \cite{Abbott2017c}, is not the optimal way to account for this; the distribution of model log-distance ratios after Bayesian Averaging is non-Gaussian and has significant density in the tails of the distribution that would not be captured by placing a lower limit on the uncertainty of the log-distance ratio or peculiar velocity in either our model (Eq.~\ref{eq:recvelpdf}) or that of \cite{Abbott2017c}.

\begin{figure*}
\centering
\includegraphics[width=0.8\textwidth]{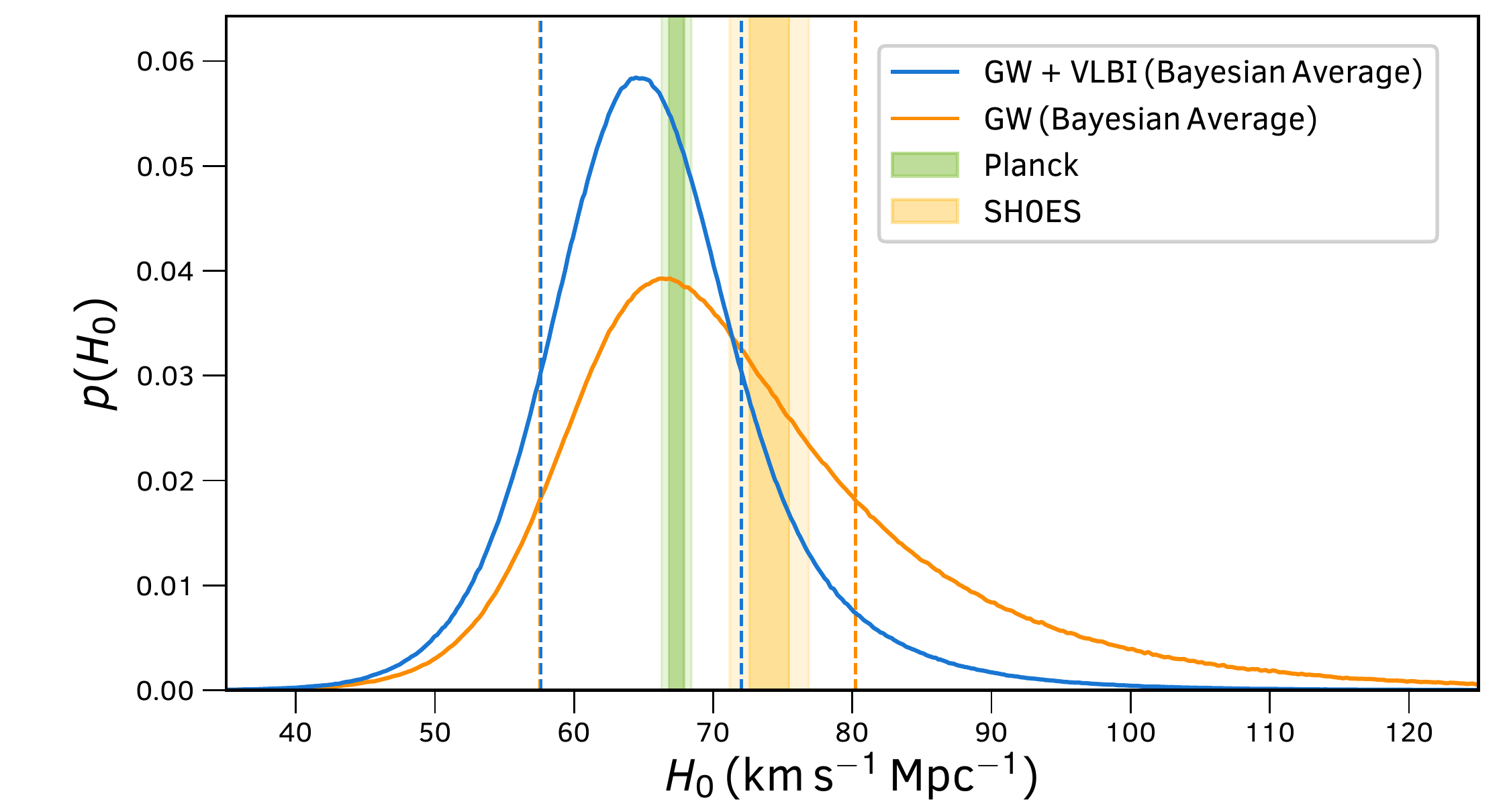}
  \caption{Posterior probability distribution functions for the Hubble constant from our reanalysis of GW170817 and the properties of its host NGC4993. The blue and orange curve show the constraints with and without the inclusion of extra information on the source inclination from observations of the radio afterglow respectively, the vertical dashed lines show the upper and lower equal likelihood bounds encapsulating $68\%$ of the posteriors. For both cases we have performed a Bayesian Average over the 154 different combinations of total observed redshift and log-distance ratio identified in this work. The vertical bars show the $1$ and $2\sigma$ bounds on the Hubble constant from the \protect\cite{Planck2018} (green) and the local distance ladder as measurement by the SH0ES collaboration (orange; \protect\citealt{Riess2019}).}
  \label{fig:comparison}
\end{figure*}

Finally, Figure~\ref{fig:comparison} shows a comparison of our new results for the Hubble constant from GW170817 with and without the inclusion of additional VLBI observations and modelling of the radio afterglow against those from the \cite{Planck2018} and the SH0ES collaboration \citep{Riess2019}. The maximum \textit{a posteriori} value of $H_{0}$ is in better agreement with the \cite{Planck2018} than \cite{Riess2019}, particularly now that our Bayesian Average has lowered this compared to the canonical constraints. However, our results also still agree with the SH0ES at just over $1\sigma$ even for the more constrained case based on the analysis of \cite{Hotokezaka2019}. This would not be the case if we had not marginalised over our uncertainty in the observed properties of NGC4993; there are a number of choices of observed redshift and peculiar velocity that, if treated the same way as the canonical model, would have led to a larger discrepancy between the constraints from GW170817 and SH0ES. For example, cases 2, 4 and 6 in Table~\ref{tab:results} would all be discrepant with SH0ES by $>2\sigma$ if treated in isolation. More completely including and accounting for uncertainty in the observed quantities is of key importance for nearby Standard Siren measurements to ensure we do not reach biased conclusions. One silver-lining is that as more local measurements are obtained, the random component of the peculiar velocity errors will average out, however this will require a considerable number of independent measurements and they will still remain susceptible to coherent systematic errors. Standard Sirens at larger distances will also be less affected by changes in the peculiar velocity correction.

\subsubsection{Comparison to other recent results}
Concurrent to this work, two other studies were produced that examined how different choices for the peculiar velocity of NGC4993 used to analyse GW170817 affected the Hubble Constant constraints. Whereas we have focused on direct estimates of galaxy peculiar velocities from secondary distance indicators, \cite{Mukherjee2019} look at reconstruction of the local velocity field, using a new algorithm that forward models the observed redshift distribution of galaxies using $N$-Body methods, which simultaneously recovers the real-space overdensity and velocity fields. Applying a newly derived velocity correction to the data from \cite{Hotokezaka2019} they arrive at $H_{0}=69.3^{+4.5}_{-4.0}\,\mathrm{km\,s^{-1}\,Mpc^{-1}}$ to be compared to our result of $H_{0}=64.8^{+7.3}_{-7.2}\,\mathrm{km\,s^{-1}\,Mpc^{-1}}$. The two results are consistent at the $1\sigma$ level, with our results reporting larger error bars and a lower value of $H_{0}$. This is not surprising given that the peculiar velocity at the location of NGC4993 reported in \cite{Mukherjee2019} is very similar to that used in the canonical analysis, whereas we find that the majority of alternative data choices give a larger peculiar velocity and hence smaller $H_{0}$. However, our method of averaging over many choices of peculiar velocity data ensures consistency between the constraints. The benefit of using the reconstructed field is that it provides a consistent and coherent interpolation of the velocity field across a wide cosmological volume where the selection functions inherent in the data can be modelled and included. An interesting avenue for future work would be to see how the reconstructed velocity field changes when the choice of data to fit against is changed (in a similar vein to what we have investigated here), and whether there are differences that arise in the recovered peculiar velocity that need to be accounted for.

The second study, \cite{Nicolaou2019} has a more sizeable overlap with our work. They investigate how the $H_{0}$ constraints change when the smoothing scale used to infer the peculiar velocity from the 6dFGSv peculiar velocity catalogue is included as a free parameter in the Bayesian model. They obtain $H_{0}=68.6^{+14.0}_{-8.5}\,\mathrm{km\,s^{-1}\,Mpc^{-1}}$ using the original \cite{Abbott2017c} data, compared to our new result of $H_{0}=66.8^{+13.4}_{-9.2}\,\mathrm{km\,s^{-1}\,Mpc^{-1}}$. Again the two results are consistent, with our study finding a slightly lower value. Both studies have accounted for variations in the smoothing scale on the recovered values of $H_{0}$, with \cite{Nicolaou2019} adopting a more rigourous marginalisation over this choice compared to our Bayesian average (which could be likened more to a grid based marginalisation). However, in our study we also account for changes in the input peculiar velocity catalogue in addition to the smoothing scale applied to the catalogue; as both 2MTF and CF3 (for small smoothing lengths) prefer larger peculiar velocities than 6dFGSv, we find a lower value for $H_{0}$. An obvious  way forward for future work would be to combine both our methodologies and adopt the smoothing scale as a free parameter in the Bayesian model for each choice of peculiar velocity catalogue \textit{and} then average over the choice of input data. This would significantly reduce the number of unique MCMC chains/evidence calculations used in our work, reducing the computational cost. 

\section{Conclusions} \label{sec:conclusion}
Standard Siren measurements of the Hubble constant have the potential to rival those from Standard Candles or Rulers. However, current measurements are limited by our knowledge of the host velocities, in particular the peculiar velocity. In this work we have demonstrated that current measurements of the observed redshift and peculiar velocity (or rather log-distance ratio) obtained using different methods and data for the single Standard Siren measurement from GW170817 and NGC4993 contain considerable uncertainty that is not, and likely cannot, be fully understood. This leads to uncertainties on the recovered Hubble constant larger than we would naively assume.

We have presented one way to account for this uncertainty using Bayesian Model Averaging and obtain constraints of $H_{0}=66.8^{+13.4}_{-9.2}\,\mathrm{km\,s^{-1}\,Mpc^{-1}}$; and $H_{0}=64.8^{+7.3}_{-7.2}\,\mathrm{km\,s^{-1}\,Mpc^{-1}}$ without and with the inclusion of high-resolution measurements of the radio counterpart. These are lower and have substantially larger errors than those originally quoted in \cite{Abbott2017c} and  \cite{Hotokezaka2019}. In the course of this work, we have also developed a model for the posterior distribution of the Hubble constant that works more closely with the observed quantities from galaxy redshift and peculiar velocity surveys. However, the main conclusion from this work is that greater understanding is needed of the limitations of current methods to obtain total and peculiar velocities for Standard Siren measurements, how these compare, and how these can be combined. This will remain an important consideration in the future as more Standard Sirens are detected, at least until we have a large enough number of measurements, or more measurements at larger observed redshifts, to mitigate the effects of peculiar velocity errors.

\section{Acknowledgements}
We thank David Parkinson for useful discussions. We are also especially grateful to Kenta Hotokezaka and the LIGO Collaboration for providing their posterior samples for GW170817 and to Alex Merson, D. Heath Jones, Matthew Colless and the authors of all the publicly available group and peculiar velocity catalogues used in this work. This research would not have been possible without these data. 

This  research  was  supported by  the  Australian  Government  through  the  Australian Research  Council’s  Laureate  Fellowship  funding  scheme (project FL180100168). This research has made use of NASA's Astrophysics Data System Bibliographic Services and the \texttt{astro-ph} pre-print archive at \url{https://arxiv.org/}. Plots in this paper were made using the {\sc matplotlib} plotting library \citep{Hunter2007} and the {\sc chainconsumer} package \citep{Hinton2016}.

\appendix
\section{Bayesian model for $H_{0}$ using peculiar velocity instead of log-distance ratio.}
In this appendix we present the posterior probability of $H_{0}$ based on the framework in Section~\ref{sec:hubble} given measurements of the object's peculiar velocity as opposed to the log-distance ratio. This may be appropriate for the case where the peculiar velocity of the group containing the host galaxy or object is obtained from reconstructions of the velocity field as opposed to direct measurements from peculiar velocity surveys such as CF3, 2MTF or 6dFGSv. The method is very similar to that originally used in \cite{Abbott2017c}, however we treat the object's peculiar velocity as the measurement to fit against, rather than the observed redshift, and we make no approximations on the relationship between the various redshifts and velocities.

We start by writing the posterior probability
\begin{align}
& p(H_{0}|x_{GW},\langle v^{\mathrm{obj}}_{\mathrm{p}} \rangle,\langle z_{\mathrm{obs}} \rangle,\langle z^{\mathrm{Sun}}_{\mathrm{p}} \rangle) \propto p(H_{0}) \int \bigg[p(x_{GW}|d_{L},\mathrm{cos}\,\iota) \notag \\
& \times  p(\langle v^{\mathrm{obj}}_{\mathrm{p}} \rangle|H_{0},d_{L},z_{\mathrm{obs}},z^{\mathrm{Sun}}_{\mathrm{p}})\,p(\langle z_{\mathrm{obs}} \rangle|z_{\mathrm{obs}})\,p(\langle z^{\mathrm{Sun}}_{\mathrm{p}} \rangle|z^{\mathrm{Sun}}_{\mathrm{p}}) \notag \\
& \times p(d_{L})p(\mathrm{cos}\,\iota)p(z_{\mathrm{obs}})p(z^{\mathrm{Sun}}_{\mathrm{p}})\bigg] \mathrm{d}d_{L}\,\mathrm{dcos}\,\iota\,\mathrm{d}z_{\mathrm{obs}}\,\mathrm{d}z^{\mathrm{Sun}}_{\mathrm{p}},
\label{eq:h0posterioralt}
\end{align} 
where $\langle v^{\mathrm{obj}}_{\mathrm{p}} \rangle$ is the measured peculiar velocity of the object and other terms are as defined in Section~\ref{sec:hubble}. If we adopt a Gaussian distribution for the object's measured peculiar velocity, 
\begin{equation}
p(\langle v^{\mathrm{obj}}_{\mathrm{p}} \rangle|H_{0},d_{L},z_{\mathrm{obs}},z^{\mathrm{Sun}}_{\mathrm{p}}) = N[v^{\mathrm{obj}}_{\mathrm{p}}(H_{0},d_{L},z_{\mathrm{obs}},z^{\mathrm{Sun}}_{\mathrm{p}}),\sigma_{v^{\mathrm{obj}}_{\mathrm{p}}}](\langle v^{\mathrm{obj}}_{\mathrm{p}} \rangle),
\end{equation}
where $\sigma_{v^{\mathrm{obj}}_{\mathrm{p}}}$ is the measurement error on the object's peculiar velocity, all that remains is for us to write the model peculiar velocity in terms of the parameters $H_{0}$, $d_{L}$, $z_{\mathrm{obs}}$ and $z^{\mathrm{Sun}}_{\mathrm{p}}$. We do this by first computing the object's peculiar redshift 
\begin{equation}
z^{\mathrm{obj}}_{\mathrm{p}} = \frac{1+z_{\mathrm{obs}}}{(1+\bar{z}(d_{L},H_{0},z_{\mathrm{obs}}))(1+z^{\mathrm{Sun}}_{\mathrm{p}})} - 1
\end{equation}
where the cosmological redshift $\bar{z}(d_{L},H_{0},z_{\mathrm{obs}})$ is computed by numerically inverting the redshift-distance relation. The object's peculiar redshift is then converted to a peculiar velocity using
\begin{equation}
v^{\mathrm{obj}}_{\mathrm{p}} = c\frac{(1+z^{\mathrm{obj}}_{\mathrm{p}})^{2}-1}{(1+z^{\mathrm{obj}}_{\mathrm{p}})^{2}+1}.
\end{equation}
Although this seem more complex than the model used in \cite{Abbott2017c}, it is not computationally demanding or difficult to implement; the numerical inversion of the redshift-distance relationship can be achieved extremely efficiently using, for instance,  spline interpolation. However, this model makes no assumptions about the relationship between the various redshifts which could bias constraints from Standard Sirens.
\end{document}